\documentstyle[prb,aps,preprint]{revtex}
\tightenlines
\author{Steffen Trimper and Simone Artz}
\address{Fachbereich Physik\\
Martin-Luther-Universit\"at\\Halle D-06099, Germany}
\title{On A Local Carnot Engine}
\begin{document}
\draft
\date{\today}
\maketitle
\begin{abstract}
Starting from a master equation in a quantum Hamilton form we study 
analytically a nonequilibrium system which is coupled locally to two heat 
bathes at different temperatures. Based on a lattice gas description an 
evolution equation for the averaged density in the presence of a temperature 
gradient is derived. Firstly, the case is analysed where a particle is removed 
from a heat bath at a fixed temperature and is traced back to the bath at another 
temperature. The stationary solution and the relaxation time is discussed. 
Secondly, a collective hopping process between different heat bathes is 
studied leading to an evolution equation which offers a bilinear coupling 
between density and temperature gradient contrary to the conventional approach. 
Whereas in case of a linear decreasing static temperature field the relaxtion 
time offers a continuous spectrum it results a discrete spectrum for a 
quadratically decreasing temperature profile.\\[1cm]

\end{abstract} 

\section{Introduction}

Within the classical equilibrium thermodynamics the Carnot cycle 
had been considered as the standard example for studying the efficiency of  
heat devices acting between two heat bathes. The analysis is restricted to 
reservoirs with fixed temperatures and furthermore to the reversible limit. 
Consequently Carnot's engine works only with a fixed rate of heat provided 
from the reservoirs.\\ 
Recently, Velasco et al \cite{vrmh} studied a finite time Carnot refrigerator 
to get an upper bound for the coefficient of performance of endoreversible 
refrigerators. They found an upper bound for the mentioned coeffcient depending 
on the ratio between the temperature of the cold and the hot reservoir. \\
There is a general interest in nonequilibrium systems with two temperatures 
\cite{cglv,br,s,sb,agr,e,rws,vs}. The analysis is motivated by searching for some 
generic features of nonequilibrium steady states. In particular, the question 
appears for a universal behavior under nonequilibrium conditions. As an example, 
a two-temperature, kinetic Ising model is investigated \cite{cglv} extended to 
a diffusive kinetic system in \cite{br}. The authors found a bicritical point 
where two nonequilibrium critical lines meet. The analysis is strongly 
supported by Monte Carlo simulations in two dimensions. Recently, a similar 
simulation has been performed studying a two-temperature lattice gas model with 
repulsive interactions \cite{s}. A complete different approach is used 
in \cite{sb} where a thermally driven ratchet is studied under periodic, 
dichotomous temperature changes. The behavior of the engine is significant 
different from a quasistatically working one. In \cite{e} the author 
used a local heat conduction operator to study the corresponding 
thermal processes observed in complex fluids. Another approach consists of the 
analysis of a cyclic working thermodynamic devise driven by an external 
applied steady flow \cite{rws}. A nonlinear oscillator coupled to various heat 
batheshad been considered as a simple toy model \cite{vs}.\\
Here we are interested in a `local Carnot engine` on a lattice gas, i.e. each 
point of a lattice will be contacted with two heat bathes at local different 
temperatures. A particle taken away from a reservoir is created at a lattice 
point $i$ where the creation rate depends on the local temperature related to 
this lattice point. In the same way a particle is annihiliated from the 
neighboring point $j$. This particle is removed to a reservoir on a 
different temperature. As the result we consider the hopping of particles 
from a lattice point to its neighboring one wheras both points are in 
contact to heat bathes on different temperatures. Alternatively, 
a model is studied where a particle at a certain lattice point is able to 
change its state, may be from spin up to spin down, however this flip process 
is organized by coupling to local bathes. As before the up and down states are 
also coupled to reservoirs at different temperatures. As a useful method to 
study such situations we apply the quantum formalism for nonequilibrium 
processes based upon spin variables.

\section{Quantum Approch to Nonequilibrium}

The analysis is based on a master equation 
\begin{equation}
\partial_tP(\vec n,t)=L^{\prime}P(\vec n,t) 
\label{ma}
\end{equation}
where $P(\vec n, t)$ is the probability that a certain configuration 
characterized by a state vector $\vec n = (n_1, n_2 \dots n_N) $ is realized 
at time $t$. 
In a lattice gas description each point is either empty or single occupied 
$n_i = 0, 1$. These numbers can be considered as the eigenvalues of the 
particle number operator. The dynamics is determined 
completely by the form of the evolution operator $L^{\prime }$, specified 
below, and the commutation relations of the Pauli-operators.   
Thus, the problem is to formulate the dynamics in such a way that this 
restrictions in the occupation number are taken into account explictly. The 
situation in mind can be analyzed in a seemingly compact form using the master 
equation in a quantum Hamilton formalism 
\cite{doi,gra,sp,gw,satr,sd,scsa,aldr,efgm}, for a recent review 
see \cite{sti,mg}. Within that approach \cite{doi} the probability 
distribution $P(\vec n,t)$ is related to a state vector $\mid F(t) \rangle$ 
in a Fock-space according to $P(\vec n,t) = \langle \vec n\mid F(t)\rangle$. 
The basic vectors $\mid \vec n \rangle$ are composed of Pauli-operators. 
Using the relation 
\begin{equation}
\mid F(t) \rangle = \sum_{n_i} P(\vec n,t) \mid \vec n \rangle
\label{fo2}
\end{equation}  
the master Eq. (\ref{ma}) can be transformed into an equivalent 
one in a Fock-space
\begin{equation}
\partial_t \mid F(t)\rangle = L \mid F(t) \rangle
\label{fo1}
\end{equation}
where the operator $L'$ in (\ref{ma}) is mapped onto the operator $L$ 
in Eq.(\ref{fo1}). It should be emphasized that the procedure is up to now 
independent on the realization of the basic vectors. Originally, the 
method had been applied for the Bose case \cite{doi,gra,pe}. Recently, 
an extension to restricted occupation numbers (two discrete orientations) 
was proposed \cite{sp,gw,satr,sd,scsa,aldr}. Further extensions to  
p--fold occupation numbers \cite{sctr2} as well as to models with kinetic 
constraints are possible \cite{sctr3}.\\
As shown by Doi \cite{doi} the average of an arbitrary physical 
quantity $B(\vec n)$ can be calculated by the average of the corresponding 
operator $B(t)$
\begin{equation}
\langle B(t) \rangle = \sum_{n_i} P(\vec n,t) B(\vec n) = 
\langle s \mid B \mid F(t) \rangle 
\label{fo3}
\end{equation} 
with the state function $\langle s \mid = \sum \langle \vec n \mid$. The 
evolution equation for an operator 
$B(t)$ reads now  
\begin{equation}
\partial_t \langle B \rangle = \langle s \mid [B(t),L] \mid F(t) \rangle
\label{kin}
\end{equation}
As the result of the procedure, all the dynamical equations govering the 
classical problem are determined by the structure of the evolution operator 
$L$ and the commutation rules of the operators. In our case the dynamics will 
be realized either by spin-flip or by exchange processes, respectively.

\section{Coupling To Heat Bathes}

The evolution operator for a local flip--process reads
\cite{sctr} 
\begin{equation}
L_i = \lambda (d^{\dagger}_i - d_i d^{\dagger}_i) + 
\gamma (d_i - d^{\dagger}_i d_i)
\label{fli1}
\end{equation}
where $\lambda$ and $\gamma$ are independent flip--rates.\\
The occupation number operator $n_i = d_i^{\dagger} d_i$ is related to the 
spin due to the relation $S_i = 1 - 2n_i$.\\
A generalization to processes under the influence of a heat bath 
with a fixed temperature $T$ is discussed in \cite{sctr}. As demonstrated in 
\cite{sctr,st1} the evolution operator has to be replaced by 
\begin{eqnarray}
L &=& \nu \sum \left[ (1 - d_i) \exp(-\beta H/2) d^{\dagger}_i 
\exp(\beta H/2) \right] \nonumber\\ 
&+& \left[ (1 - d^{\dagger}_i) \exp(-\beta H/2) d_i \exp(\beta H/2) \right] 
\label{fli4}
\end{eqnarray}
where $\nu$ is a the flip-rate defined on a microscopic time scale; 
$\beta = T^{-1}$ is the inverse temperature of the heat bath and $H$ is the 
Hamiltonian responsible for the static interaction.\\ 

\subsection{Flip-dynamics}
Whereas by Eq.(\ref{fli4}) the coupling to a global heat bath is realized we 
discuss now a further generalization by introducing two local heat bathes with 
different temperatures $T$ and $T'$, respectively. The two 
reservoirs are coupled directly to each lattice point. This situation 
can be described by an evolution operators 
\begin{equation}
L_{(f)} = \nu \sum_i \left( (1-d^{\dagger}_i) e^{-\mu n_i/2T'} d_i 
e^{\mu n_i/2T} + (1 - d_i) e^{-\mu n_i/2T'} d^{\dagger}_i 
e^{\mu n_i/2T} \right)
\label{lo1}
\end{equation}
Here $\mu$ is a characteristic energy which is necessary 
to remove a particle from the heat bath or to give it back to the bath. The 
approach reminds of using a grand canonical ensemble in equilibrium statistics. 
Therefore, the quantity $\mu$ plays the role of the chemical potential 
assumed to be identically for both processes under consideration. The 
potential $\mu$  can be positive or negative. Taking into account that the 
occupation operator $n_i$ has the eigenvalues $0$ or $1$ we get 
\begin{equation}
e^{-\mu n_i/2T'} d_i e^{\mu n_i/2T} = d_i e^{\mu/2T} \quad\quad
e^{-\mu n_i/2T'} d^{\dagger}_i e^{\mu n_i/2T} = d^{\dagger}_i e^{-\mu/2T'}
\end{equation}
Thus, the operator $d_i$ annihiliates a particles at the temperature $T$ independently  
on the temperature $T'$ of the other bath. Contrary, the operator 
$d^{\dagger}_i$ creates a particles at the temperature $T'$. The  
evolution operator $L_{(f)}$ describes the process of annihiliation and 
creation of particles within the system at different temperatures. 
Using the Eq.(\ref{kin}) and the algebraic properties of 
Pauli--operators, the evolution equation for the averaged densitiy reads
\begin{equation}
\nu^{-1} \partial_t \langle n_i \rangle = 
\exp(-\mu/(2T')) \langle 1 - n_i \rangle 
- \exp(\mu/(2T)) \langle n_i \rangle  
\label{e1}
\end{equation}
This equation can be solved easily. It results a stationary state at an 
effective temperature $T_e$
\begin{equation}
\langle n \rangle_{s} = \frac{1}{1 + \exp(\mu/T_e)} \quad\mbox{with}\quad
\frac{1}{T_e} = \frac{1}{2} (\frac{1}{T} + \frac{1}{T'})
\label{e1a}
\end{equation}
In a spin representation we obtain
\begin{equation}
\langle S \rangle_s = \frac{e^{\mu/2T} -  e^{-\mu/2T'}}{e^{\mu/2T} + e^{-\mu/2T'}} 
\label{e2}
\end{equation}
In the special case that $T = T'$ the stationary solution coincides with the 
conventional equilibrium solution
\begin{equation}
\langle S \rangle_s = \tanh\frac{\mu}{2T} \quad\quad 
\langle n \rangle_s = \frac{1}{e^{\mu/T} + 1}
\label{eq}
\end{equation}
If the temperature of one of the heat bathes tends to infinity (for instance 
$T' \to \infty$) the stationary solution is 
$$
\langle S \rangle_s = \tanh\frac{\mu}{4T}
$$
When both temperatures $T$ and $T'$ are infinitesimal different from each other 
$T' = T + \Delta T$  
the averaged occupation number is
\begin{equation}
\langle n \rangle_s = \frac{1}{e^{\mu/T} + 1} + \frac{\mu (T' -T)}{4 T^2} 
\tanh\frac{\mu}{2T}[\frac{1}{e^{\mu/T} + 1} + \frac{1}{e^{\mu/T} -1}]
\label{eq1}
\end{equation}
The Fermi--distribution as the equilibrium solution is modified in lowest 
order in $\Delta T$ by an additonal term proportional to the 
Bose--distribution.\\
The relaxation time $\tau$ related to the Eq.(\ref{e1}) is simply given by 
\begin{equation}
(\nu \tau)^{-1} = \exp(\frac{\mu}{2T}) + \exp(-\frac{\mu}{2T'})
\end{equation}
The relaxation time for $T' \ne T´$ is either enhanced for $T' < T$ or 
diminished in the opposite case.\\

\subsection{Exchange Process}
Up to now we have analysed the case of independent flip processes 
(annihiliation-creation-processes) at different temperatures without an 
internal coupling between the active particles. In the following, 
we discuss the situation that the particles can exchange their mutual 
position; with other words hopping processes are allowed between neigbored 
sites under the influence of the coupling to local heat bathes. The evolution 
operator reads 
\begin{equation}
L_{ex} = \nu \sum_{<ij>} \left[ (1 - d_i d^{\dagger}_j) 
e^{-\frac{\mu}{2}[\frac{n_i}{T_i} + \frac{n_j}{T_j}]} d^{\dagger}_i d_j 
e^{\frac{\mu}{2}[\frac{n_i}{T_i} + \frac{n_j}{T_j}]} \right]
\label{ex}
\end{equation}
It describes the exchange process between two adjacent neighboring sites, where 
the lattice site $i$ is coupled to the bath at the temperature $T_i$ and the 
site $j$ is related to $T_j$, repectively. The evolution equation for the 
averaged density can be written in the form
\begin {eqnarray}
\nu^{-1} \partial_t \langle n_r \rangle &=& \sum_{j(r)} \langle n_j \rangle
\exp[-(\mu/2)(\frac{1}{T_r} - \frac{1}{T_j})] - \langle n_r \rangle
\exp[- (\mu/2)(\frac{1}{T_j} - \frac{1}{T_r})] \nonumber\\
&-& 2 \langle n_r n_j \rangle \sinh[(\mu/2)(\frac{1}{T_j} - \frac{1}{T_r})]
\label{ex1}
\end{eqnarray}
In the special case of fixed temperatures $T_j = T_r = T$ the last equation 
is reduced to the conventional diffusion equation in a discrete representation. 
Here, the case will be studied assuming a small temperature gradient. Moreover, 
Eq.(\ref{ex1}) is investigated in the continuous limit leading to the evolution 
equation for the density $n(\vec x, t) = \langle n_r(t) \rangle l^{-d}$ 
(we set the lattice size $l =1$)
\begin{equation}
\nu^{-1} \partial_t n = \nabla^2 n + \mu n \nabla^2 \frac{1}{T} + 
\mu \left( \nabla n \cdot \nabla (\frac{1}{T}) \right) 
\label{ex2}
\end{equation}
To derive this equation we have neglected the bilinear terms in Eq.(\ref{ex1}) 
which give only rise to higher order corrections in the density. Due to the 
conservation of the spins within the exchange model the evolution equation can 
be rewritten as a continuous equation with the current $\vec j(\vec x,t)$
\begin{equation}
\vec j = -\nu \nabla n - \nu \mu n \nabla(1/T) 
\label{ex3}
\end{equation}
In contrast to the conventional nonequilibrium thermodynamics a bilinear 
coupling between the density and the temperature gradient is included in 
the current. Such a nonlinear coupling may change the physical behavior.\\ 
Using natural boundary conditions the stationary solution is
\begin{equation}
n(\vec x) = n_0 \exp(-\frac{\mu}{T(\vec x)})
\end{equation}
That means, the local density is determined by the local temperature in 
accordance with the local temperature attached to each lattice site.\\
Let us consider the special case assuming that  
$$
\nabla (\frac{1}{T}) = 2 \vec c
$$
where $\vec c$ is a constant vector leading to a decreasing temperature profile
\begin{equation}
T(\vec x) = \frac{T_0}{ 2 T_0 (\vec c \cdot \vec x) + 1}
\end{equation}
where $T_0$ is an arbitrary initial temperature. Eq.(\ref{ex2}) can be solved 
making the ansatz 
$n(\vec x,t) = \Phi(\vec x) \psi(\vec x,t)$. Chosing 
$$
\Phi = \Phi_0 \exp[-\mu (\vec c \cdot \vec x)] 
$$
$\psi(\vec x,t)$ obeys  
\begin{equation}
\dot{\psi} = - [ - \nu \nabla^2 + \nu (\mu \vec c)^2 ] \psi
\label{ham}
\end{equation}
which is nothing else as the Schr\"{o}dinger equation for a free particle. 
The relaxation time is the inverse eigenvalue of the 
Hamiltonian $\hat{H} = - \nu \nabla^2 + \nu (\mu \vec c)^2$ 
defined in Eq.(\ref{ham}): 
\begin{equation}
\tau_k = \frac{1}{\nu (k^2 + (\mu \vec c)^2 )} = 
\frac{1}{\nu \left( k^2 + \mu^2 (\nabla(1/T))^2/4 \right) }
\end{equation}
where $\vec k$ is the wave vector. There is a gap in the quasi-continuous 
relaxation spectrum for $\vec k \to 0$. Moreover, the relaxation time depends 
on the temperature gradient.
The solution for the density is 
\begin{equation}
n(\vec x, t) = n_0 \exp \left[ ( - \mu \vec c + i \vec k )\cdot \vec x - \frac{t}{\tau_k} \right]
\end{equation}
Let us consider the further solvable case
$$
\nabla(\frac{1}{T}) = 2 b \vec x
$$ 
which leads to a quadratically decreasing temperature profile, b is a constant. 
$$
T(\vec x) = \frac{T_0}{T_0 b \vec x^2 + 1}
$$
The same procedure as used before yields the Hamiltonian $\hat{H}$ of the 
d-dimensional harmonic oszillator
$$
\hat{H} = - \nu \nabla^2 + \nu (\mu b)^2 \vec x^2 - \nu \mu b d
$$
It results a discrete relaxation time spectrum where the ground state energy 
of the harmonic oscillator is cancelled due to the last term in $\hat{H}$.
\begin{equation}
\tau = \frac{1}{2 \nu \mu b (m_1 + m_2 + \ldots + m_d)}  
\end{equation} 
where the $m_i$ are integer numbers.
Obviously, the analysis can be extended to other static temperature profiles 
such as an arbitrary radial symmetric one without changing the results 
substantially.\\
A temperature flow is allowed if the heat bathes are coupled. This process 
leads to an equalization of temperatures. Assuming that this 
process follows the conventional heat conduction equation 
$$
T(\vec x, t) = \frac{1}{(4 \pi \lambda t)^{d/2}} 
\exp(-\frac{\vec x^2}{4 \lambda t}) \qquad \lambda \quad \mbox{heat conduction} 
$$
we conclude that Eq.(\ref{ex2}) can also be transformed into a 
Schr\"odinger-like equation
\begin{equation}
\partial_t \psi(\vec x,t) = - [- \nu \nabla^2  + V(\vec x, t) ] \psi(\vec x,t)
\end{equation}
where the potential $V(\vec x, t)$ is known. The procedure yields a similar 
result for an arbitrary temperature distribution of the form 
$T(\vec x,t) = t^{-\alpha} g(x^2/t)$.

\section{Conclusions}
\noindent Here we have considered a generalization of the well known Carnot 
cycle with open flow. Each point of a lattice is related to a local heat bath 
hold on different temperatures. Introducing the conventional temperature 
means that the system is not too far from equilibrium. Consistently with this 
assumption is the consideration of small gradients in temperature leading to 
a heat transport. Such a temperature gradient is coupled to the creation and 
annihiliation of particles or to an exchange process where a particle is 
created at a certain lattice point at a fixed temperature and annihiliated at 
another point with another but fixed temperature. Using a quantum formalism 
for the master equation we can derive an evolution 
equation for the density which offers already in a mean field like approximation 
a bilinear coupling between density and temperature gradients. This leads to 
a stationary state where the local density is related to a local temperature. 
The dynamics is studied for some special cases with a static temperature 
profile. The relaxation time spectrum offers a different behavior depending 
on the realization of the temperture field. Moreover, the relation time depnds 
also on the temperature field.


\end{document}